\documentclass[11pt,oneside]{article}
\usepackage{graphicx}
\begin{document}
 
\title{{\bf Evolution of the Velocity Ellipsoids in the Thin Disk of 
the Galaxy and the Radial Migration of Stars}}
\author{{\bf V.~V.~Koval', V.~A.~Marsakov, T.~V.~Borkova}\\
Institute of Physics, Southern Federal University,\\
Rostov-on-Don, Russia\\
e-mail: koval@ip.rsu.ru, marsakov@ip.rsu.ru, borkova@ip.rsu.ru}
\date{accepted \ 2009, Astronomy Reports, Vol. 86 No. 9, P.1117-1126}
\maketitle

\begin {abstract}
Data from the revised Geneva--Copenhagen catalog are used to study the 
influence of radial migration of stars on the age dependences of 
parameters of the velocity ellipsoids for nearby stars in the thin
disk of the Galaxy, assuming that the mean radii of the stellar orbits 
remain constant. It is demonstrated that precisely the radial migration 
of stars, together with the negative metallicity gradient in the thin 
disk,are responsible for the observed negative correlation between the 
metallicities and angular momenta of nearby stars, while the angular 
momenta of stars that were born at the same Galactocentric distances do
not depend on either age or metallicity. The velocity components of the 
Sun relative to the Local Standard of Rest derived using data for stars 
born at the solar Galactocentric distance are 
$(U_{\odot}, V_{\odot},W_{\odot})_{LSR} =(5.1\pm0.4, 7.9\pm0.5, 7.7\pm0.2)$
~km\,s$^{-1}$. The two coordinates of the apex of the solar motion 
remain equal to $\langle l_{\odot} \rangle = 70^{\circ} \pm 7^{\circ}$ 
and $\langle b_{\odot} \rangle = 41^{\circ} \pm 2^{\circ}$, within the 
errors. The indices for the power-law age dependences of the major,
middle, and minor semi-axes become $0.26\pm0.04$, $0.32\pm0.03$, 
and $0.07\pm0.03$, respectively.As a result, with age, the velocity 
ellipsoid for thin-disk stars born at the solar Galactocentric distance 
increases only in the plane of the disk, remaining virtually constant 
in the perpendicular direction. Its shape remains far from equilibrium, 
and the direction of the major axis does not change with age: the 
ellipsoid vertex deviation remains constant and equal to zero within 
the errors $(\langle L\rangle = 0.7^{\circ} \pm 0.6^{\circ})$, 
$(\langle B\rangle = 1.9^{\circ} \pm 1.1^{\circ})$. Such a small 
increase in the velocity dispersion perpendicular to the Galactic plane 
with age can probably be explained by ''heating'' of the stellar system 
purely by spiral density waves, without a contribution from giant 
molecular clouds.

\end {abstract}

\section*{Introduction}

This study is a logical continuation of our earlier
paper~[1], where we used data from the revised
Geneva--Copenhagen catalog~[2] to investigate the
age and metallicity dependences of parameters of the
velocity ellipsoids for selected stars of the thin disk
of the Galaxy. In particular, we showed that the velocity
ellipsoid for solar-neighborhood thin-disk stars
increases with age, only slightly circularizes its shape,
turns slightly toward the direction of the Galactic
center, and loses angular momentum. This behavior
is in full agreement with the idea that the stellar velocity
dispersions increase with age under the action
of various relaxation processes, while the subsystem
itself approaches an equilibrium state. However, it
was simultaneously demonstrated that the velocity
ellipsoid for stars of mixed age increased with increasing
metallicity, while displaying a weak tendency to
increase its sphericity and turn toward the direction
of the Galactic center. Thus, the velocity behaved in
the same way as it did with variations in age; however
the rotational velocity around the Galactic center
monotonically increased with decreasing metallicity
(!), rather than decreasing. This last effect was first
described in~[3], where it was discovered in an analysis
of the mean tangential velocities and metallicities
of nearby stars, and was called the ''kinematics--
metallicity paradox for thin-disk Galactic stars.'' In~[1], 
we suggested that the metallicity dependences for
the velocity ellipsoids inferred for solar-neighborhood,
thin-disk stars could be due to the radial migration of
stars.

The existence of radial migration of stars was first
proposed by Grenon~[4], in order to explain certain
properties of neaby stars. (Since we are considering
here not systematic variations in the positions
of stars, but only fluctuations in their positions in
non-circular orbits, it would be more correct to call
these radial motions of stars, however, we have retained
the established term from the literature.) According
to modern concepts, the orbits of stars born
from interstellar material moving in circular orbits increase
their eccentricity with time due to interactions
with perturbations in the gravitational potential of the
Galaxy. However, it is believed that the mean radii
of their orbits should remain essentially constant in
this process, reflecting the Galactocentric distance of
their birth~[5,\,6]. As a result, some fraction of stars
currently located near the Sun have migrated there
from other Galactocentric distances~[7], and this can
distort the parameters of the velocity ellipsoids for
nearby stars. If we suppose that stars of any metallicity
in the thin disk of the Galaxy form from interstellar
material that moves in circular orbits, the orbits for
stars of different metallicities should, on average, vary
in roughly the same way with age. However, it turned
out that the semi-axes of the velocity ellipsoids for
nearby stars systematically (although only slightly)
increased with decreasing metallicity, while the deviation
of the vertex—the direction of the apex of
the solar motion--and the angular momentum of the
stars varied quite strongly with metallicity~[1, Fig. 4].
It appears that precisely the radial migration of stars
could lead to metallicity dependences for the parameters
of the velocity ellipsoids for nearby stars, since
there is a negative radial metallicity gradient in the
thin disk (see, for example,~[8]).

The current study is concerned with a detailed
verification of the influence of the radial migration
of stars on the parameters of the velocity ellipsoids
for nearby stars, using the same sample as in our
previous work.

\section*{OBSERVATIONAL DATA.}

We will now briefly summarize the main aspects
of how our sample of stars was constructed. We
used data from the new version of the Geneva--
Copenhagen survey [2], which contains the ages,
metallicities, spatial-velocity components, and
Galactic-orbit elements for $\approx14 000$ F-\,G dwarfs
and subgiants brighter than $V\approx 8.5^{m}$. (This same
group continued the process of refining the stellar
parameters in the catalog (see~[9]), but the resulting
changes in the distances, temperatures, and ages of
the stars were smaller than in the first revision, and,
according to those authors, the features of the corresponding
age--metallicity dependences remained
virtually the same. The newest version of the catalog
is not yet available.) We first eliminated binary stars,
very evolved stars $\delta M_{V} > 3^{m}$, and stars with uncertain
ages ($\epsilon_{ t} > \pm 3$ billion years) from the sample.
Moreover, to avoid boundary effects, we restricted
our range of effective temperatures to 5200--7000 K.
Thin-disk stars were selected according to our modification
of the method~[1], first proposed in~[10], which
calculates the probability of membership of stars to
the thin-disk and thick-disk subsystems based on
their spatial-velocity components. We determined
these probabilities using our refined values~[1] of the
Geneva--Copenhagen survey data for the dispersions
of each of the three spatial-velocity components
($\sigma_{U}, \sigma_{V}, \sigma_{W}$), the mean rotational 
velocities ($V_{rot}$),and the relative populations of stars in the 
two disk subsystems at the solar Galactocentric distance. Our
kinematic criterion proved to be in good agreement
with the chemical compositions of the stars: using
this criterion naturally minimizes the numbers of
stars with high relative abundances of magnesium
in the thin disk and with low relative abundances of
magnesium in the thick disk. Further, we removed
members of the most numerous moving groups from
the sample: Sirius, the Hyades, the Pleiades, Coma
Berenices, and $\xi$\,Hercules. The resulting sample
contains 4549~single F--G dwarfs and subgiants of
the thin disk of the Galaxy lying within $\approx 150$\,pc of the
Sun.

\section*{DEPENDENCE OF THE VELOCITY ELLIPSOIDS FOR STARS OF MIXED AGES 
AND METALLICITIES ON THEIR MEAN ORBITAL RADII.}

Let us first investigate how the velocity ellipsoids
for stars of mixed ages and metallicities depend on
the mean radii of their orbits. (The mean orbital radius
is $R_{m} = (R_{a} + R_{p})/2$, where $R_{a}$ and $R_{p}$ 
are the apogalactic and perigalactic orbital radii; we take 
the distance of the Sun from the Galactic center to be
$R_{0} = 8.0$\,kpc.)We separated our sample into 12 
subgroups with equal numbers of stars in narrow bands
of $R_{m}$ and calculated the parameters of the velocity
ellipsoids for each subgroup.We calculated the velocity
ellipsoids and the solar velocity relative to the local
centroids using the formulas of Ogorodnikov~[11], and
the errors in these parameters using the formulas of
Parenago~[12]. The corresponding dependences are
shown in Fig.~1. The first two plots in this figure show
that neither the semi-axes nor their ratios depend significantly
on $R_{m}$: none of these correlations are significant
($P > 5\,\%$). The probabilities that the correlations
in Fig.~1c arose by chance, $P_{L} \ll 1\,\%$ and 
$P_{B} > 5\,\%$, indicate that the $R_{m}$ dependence of the vertex
deviations in longitude is highly significant, while
the dependence of the vertex deviations in latitude
is not significant. Although the vertex coordinates
display a clear dependence on $R_{m}$, their deviations
from zero nowhere exceed $3^{\circ}$; their mean values are
$(\langle L\rangle = 0.7^{\circ} \pm 0.4^{\circ})$, 
$(\langle B\rangle = 0.4^{\circ} \pm 0.3^{\circ})$, indicating that
these deviations are insignificant. We checked this
result by dividing the entire sample into four groups
in metallicity and constructing analogous plots for
each group. The slopes of the $R_{m}$ dependences for
the two coordinates of the vertex deviations appear
random, and the subgroups of stars with different
metallicities do not show any systematic behavior.
Thus, we conclude that the vertex deviations for stars
with different $R_{m}$ values are the same and equal to
zero within the errors.

At the same time, the integrated indices for the
ellipsoids demonstrate quite significant dependences.
The essentially linear and very substantial decrease in
the V component of the solar velocity with increasing
$R_{m}$ ($P \ll 1\,\%$) is especially striking. The dependence
of the U component on $R_{m}$ is also highly significant,
though its slope is appreciably smaller. Since the
solar motion reflects the motion of the corresponding
centroids, this means that the angular momentum
of the stars decreases with the mean radii of their
orbits. This makes sense: stars located closer to the
apogalactic radius of their orbits rotate about the
Galactic center more slowly at a given Galactocentric
distance. The $U$ component also decreases here.
Since we are dealing with a statistical ensemble and
the $U$ and $V$ components vary differently, the direction
toward the apex of the solar motion also depends on
the orbital phase at which the majority of stars in each
$R_{m}$ subgroupare located. Both the apex-coordinate
dependences are highly significant ($P \ll 1\,\%$), but are
not shown in Fig.~1 because they are non-linear: the
longitude of the apex is roughly constant for small
orbits, then begins to grow sharply after crossing
the mean orbital radius for the solar Galactocentric
distance. The apex latitude also begins to deviate
sharply from a constant value near this radius. This
behavior is due to the fact that, near the solar orbital
radius, all three components of the solar velocity are
roughly equal to each other in magnitude, while the
ratios of their values reverse with further increase in
$R_{m}$.We checked the behavior of the $R_{m}$ dependences
of all ellipsoid parameters by dividing the sample into
four metallicity groups. These all demonstrate virtually
identical dependences for all parameters. Even
the direction of the vertex in latitude exhibits similar
fluctuations at high values of the mean orbital radii in
all the metallicity groups.

Using our linear regression fits for each component
of the solar motion in Fig.~1d, we can determine
the components of the solar motion relative to
the local centroid (i.\,e., the Local Standard of Rest,
LST) for stars that were born at the Galactocentric
distance corresponding to the current position of the
Sun. These were found to be 
$(U_{\odot}, V_{\odot},W_{\odot})_{LSR} =(5.1\pm0.4, 7.9\pm0.5, 7.7\pm0.2)$~km\,s$^{-1}$.

Thus, we see that the tangential component of the
solar velocity relative to the corresponding centroid,
and, as a consequence, the longitude of the apex of
the solar motion, depend most strongly on the mean
orbital radius, while the deviation of the major semiaxes
of the ellipsoid from the direction toward the
Galactic center does not vary significantly. Thus, the
angular momenta of stars located in the solar neighborhood
are directly dependent on the mean radii
of their orbits. Let us now consider the metallicity
dependences for these two most significant ellipsoid
parameters in Fig.~1 for stars with narrow ranges of
mean orbital radii.

\section*{METALLICITY DEPENDENCE FOR THE APEX OF THE SOLAR MOTION 
AND ANGULAR MOMENTUM FOR STARS WITH DIFFERENT MEAN ORBITAL RADII}

To construct the indicated dependences, we divided
the initial sample into four groups with roughly
equal numbers of stars separated by $R_{m}$ values of
7.37, 7.72, and 8.10\,kpc, then divided each into seven
subgroups in metallicity. The left and right plots in
Fig.~2 present the metallicity dependences of $V$ and
the longitude of the apex of the solar motion relative
to the corresponding centroids for each interval of
$R_{m}$. The dependences of the angular momentum are
virtually flat for all orbital radii. [Although significant
correlations are found in some cases (see the captions
next to the dependences), their slopes vary chaotically
with variations in the mean orbital radii of the stars,
testifying that these dependences are due to selection
effects associated with the appreciable widths of the
$R_{m}$ intervals.] With regard to the longitude of the
apex, in contrast to the total sample of thin-disk stars,
for which we observe an appreciable growth in $L$
with increasing metallicity~[1, Fig.~4c], we can see
everywhere in Fig.~2b clear dependences showing the
opposite tendency. In all intervals with $R_{m} < R_{0}$, the
dependences are essentially horizontal, and only in
the intervals with the largest mean orbital radii are
the slopes substantial, demonstrating the large width
of the total interval considered. It is striking that, for
the subgroups of any metallicity, the longitude of the
apex of the solar motion is oriented in the opposite
direction for stars with larger and smallermean orbital
radii. This can be understood if the angular momenta
of stars with larger orbits are larger at the solar
Galactocentric distance, since they are located further
from their apogalactic distances, where the tangential
velocity component is minimum.

We also traced the metallicity dependences of the
other parameters of the velocity ellipsoids for stars in
narrow intervals of $R_{m}$ (not shown in the figure in
order to conserve space). It turned out that the dependences
for parameters of a given type do not depend
on the size of the stellar orbits. Only the semi-axes
increase (slightly) with decreasing metallicity within
each groupin $R_{m}$. However, this can be explained by
the fairly large widths of the intervals in $R_{m}$, especially
in the group with the smallest orbital radii.

Thus, we conclude that no metallicity dependence
is observed for the velocity ellipsoids of stars having
the same mean orbital radii. Let us now consider how
the velocity ellipsoids for stars born within narrow
ranges of Galactocentric distance depend on age, and
whether they will display such substantial variations
of the angular momentum with age.

\section*{AGE DEPENDENCES OF THE VELOCITY ELLIPSOIDS FOR STARS WITH VARIOUS
MEAN ORBITAL RADII}

To construct these dependences, we divided the
original sample into four groups in $R_{m}$ with roughly
equal numbers of stars separated by 7.37, 7.72, and
8.10\,kpc, and then divided each of these into 12 subgroups
in age. Since the defining parameters of the
velocity ellipsoids are their semi-axes and the components
of the solar motion, we present here only these,
and only briefly describe the behavior of the secondary
parameters. The first row of plots in Fig.~3 presents
the age dependences of the ellipsoid semi-axes $\sigma_{i}(t)$
for all four groups. All the dependences were fit with
power laws of the form $\sigma_{i}\sim(t)^{\gamma}$, 
where $\sigma_{i}$ is the dispersion of the corresponding 
velocity component and $t$ the ages of the stars. The 
corresponding power-law indices are indicated in the figure 
next to each curve. The correlation coefficients are not 
presented,since all the dependences are highly significant 
($P \ll 1\,\%$). Let us consider the variations of the power-law
indices for a particular type of dependence in all the
plots. For example, the power-law index for the semi-major
axis displays a weak tendency to increase with
increasing mean stellar radius: the values of $\gamma_{1}$ for the
first two groups are significantly smaller than those at
the right-hand side of the plots. The power-law index
for the second semi-axis does not display any systematic
trend, but the contrast between the high values
for the two middle groups and small values for the two
end groups is striking. We can see that the difference
in the power-law indices came about due to the large
semi-axes of the ellipsoids for the subgroups with
small ages in the end plots, while the semi-axes are
roughly the same size for the subgroups with larger
ages, for all intervals of $R_{m}$. The origin of this is the
absence of stars with unperturbed initial orbits in the
two end groups, due to selection effects. In contrast to
the middle semi-axes, the minor semi-axes demonstrate
very small and roughly equal power-law indices
for both groups with intermediate mean orbital radii,
while both end groups have $\gamma_{3}$ values roughly twice
as high, with this difference lying beyond the errors
(see captions on the plots). Here, the semi-axes for
the end groups were found to be larger than those for
the middle groups due to the same selection effects,
for the entire age interval. Overall, we can see that
the power-law indices for the age dependences of the
ellipsoid semi-axes also depend on the mean orbital
radii of the stars and the sizes of the $R_{m}$ intervals;
therefore, to derive parameters of the velocity ellipsoids
undistorted by selection effects, we must select
only stars with mean orbital radii that are close to the
current solar Galactocentric distance.

The semi-axis ratios in all $R_{m}$ groups are essentially
independent of age, but the sequence of ratios
of the middle to the major and minor to the major
further from each other than those for the two end
groups, where the influence of selection effects is
manifest. All the age dependences of the vertex deviations
have only low significance ($P \geq 5\,\%$), and
sometimes change their sign between neighboring
subgroups, testifying that their origin is most likely
selection effects.

The bottom row of plots in Fig.~3 presents the
age dependences for the integrated parameters of the
ellipsoids for the same groups of stars in narrow intervals
of the mean orbital radii. The dependences
for the $U$ component of the solar velocity relative to
the corresponding centroids do not display systematic
variations in their slopes. Moreover, correlations
with low significance ($P > 5\,\%$) are observed in three
plots, while the substantial slope of the dependence in
Fig.~3g is due exclusively to one distant point in the
plot, and the variation of the $U$ component occur here
in the vicinity of zero velocity. The slope for the age
dependence of the $V$ component displays a systematic
decrease with increasing mean orbital radius (which
becomes even more negative for the group with the
largest orbits).An appreciable variation of the angular
momentum with age is observed only in the group
with $R_{m} < 7.37$, and is due to the fact that the mean
orbital radii in this group differ substantially from the
solar Galactocentric distance and are contained in a
wide range, so that only stars with large orbital eccentricities
reach the solar neighborhood. The slopes
in other groups are negligibly small. The systematic
nature of the behavior of the slopes testifies that the
mean angular momentum for a general collection of
stars with strictly equal mean orbital radii should
not depend on the age. The figure also shows that
the slopes for the $W$ component are small, vary sign
chaotically, and have low significance. The slopes for
the age dependences of both coordinates of the apex
likewise are small and statistically insignificant. In
summary, we can conclude that all three components
of the solar velocity relative to the centroids for stars
with mixed metallicity and age are roughly the same
for stars that are all born at the same distances from
the Galactic center.

Since the age dependences of the ellipsoids in
Fig.~3 are not always determined with certainty,
we verified these results by shifting the boundaries
between the groups several times, right to 
$\Delta R_{m} = \pm0.3$, and again constructing the dependences. 
Our conclusions above are consistent with all these
additional relations. In particular, the insignificant,
small negative slope of the $V$ component for stars
with larger orbits (i.\,e., in the open $R_{m}$ interval)
systematically increases with the distance of the orbit
for the lower boundary of the range from the solar
radius.

Thus, nearby stars with mixed metallicity demonstrate
different age dependences for their velocity ellipsoids
for different ranges of mean orbital radii. This
can be fully explained as an increase in the sample of
a deficit of stars with orbits that are close to circular
when the difference $|R_{\odot} - R_{m}|$ is increased for the
sample. Let us now investigate how these dependences
behave for stars with different metallicities.

\section*{AGE DEPENDENCE OF THE VELOCITY ELLIPSOIDS FOR STARS OF VARIOUS
METALLICITIES BORN AT VARIOUS GALACTOCENTRIC DISTANCES}

We have already described the systematic variations
of the velocity ellipsoids with metallicity, and
here consider only the variations of the ages dependences
for stars of different metallicities with $R_{m}$. To
construct Fig.~4, we divided the sample stars into two
groups with equal numbers, divided by the metallicity
$[Fe/H] = -0.13$ (which corresponds to the maximum
of the metallicity distribution of the disk stars in
our sample), then subdivided each of these into two
groups with different mean orbital radii, separated
by $R_{m} = 7.70$ (which is roughly the position of the
maximum in the $R_{m}$ distribution for disk stars). The
low-metallicity group that was closest to the Galactic
center (Fig.~3a) and the high-metallicity group that
was furthest from the Galactic center (Fig.~3d) contained
fewer stars due to the negative metallicity gradient
in the thin disk. The figure shows the behavior of
only certain parameters of the ellipsoids, namely, the
semi-axes and the components of the solar motion.
In spite of the relatively low number of stars in the
groups and the large uncertainties in the coefficients
for the linear fits in ($log \sigma_{i}$, $log t$) coordinates, 
we can trace certain tendencies in the power-law indices for
the age dependences for stars with different metallicities
in the first row of plots in Fig.~4. In particular, we
observe significantly higher power-law indices for the
semi-major axes for the low-metallicity group with
large orbits than for the corresponding group with
small orbits. This arose because the youngest stars in
the low-metallicity group that is close to the Galactic
center have larger semi-major axes than does the
distant low-metallicity group. High-metallicity stars
also display a similar difference in their power-law
indices, although this difference formally lies within
the errors. Here, this difference arose due to the
small semi-axes of the subgoup with large ages in
the near high-metallicity group. We also note higher
power-law indices for the middle semi-axes for the
subgroupwith larger orbital radii, with this difference
lying beyond the errors for the high-metallicity stars.
Both distant groups demonstrate significantly lower
power-law indices for the semi-minor axes than the
two close groups. Any possible differences in the age
dependences for the ratios of the semi-axes and the
vertex directions (not presented in the figure) are
dwarfed by the errors. We suggest that the differences
in the dispersions for the subgroups of different
ages with different mean orbital radii arise purely due
to selection effects, which affect stars with different
metallicities somewhat differently, but lead to similar
results.

The second row of plots in Fig.~4 shows that the
age dependences of the $U$ and $W$ velocity components
relative to the corresponding centroids do not display
systematic differences. Moreover, the probability that
these apparent correlations arose by chance is $P \gg 5\,\%$.
At the same time, the tangential components
of the velocity in both close subgroups turned out
to be strongly age dependent (for both, $P \ll 1\,\%$)
and nearly identical. The age dependences of these
components for the groups of stars with large $R_{m}$ are
not significant; the metal-rich group even displays a
slightly negative slope.

Thus, the stars of different metallicity demonstrate
the same variations for the age dependences of the velocity
ellipsoids with increasing orbital radius, within
the errors.

\section*{AGE DEPENDENCES OF THE VELOCITY ELLIPSOIDS FOR STARS BORN AT THE
SOLAR GALACTOCENTRIC DISTANCE}

Thus, our results provide evidence that the radial
migration of stars distorts the age dependences of
the velocity ellipsoids of stars currently located near
the Sun. Therefore, it is more correct to investigate
the age dependences for stars with approximately the
same mean orbital radii that were born at the solar
Galactocentric distance. For this, we restricted our
initial sample to single, nearby, thin-disk stars with
mean orbital radii $7.7< R_{m} <8.4$\,kpc, since the
catalog distance of the Sun from the Galactic center
is taken to be 8.0\,kpc. The resulting sample contains
1853~stars. Selection effects acting against stars with
circular orbits are virtually absent from the sample,
although there is likely some deficit of stars with orbits
that are strongly elongated and go far from the Galactic
plane, since stars in such orbits will spend a large
fraction of their time near the most distant points in
the orbits. To construct Fig.~5 and follow the age
dependences of the velocity ellipsoids, we divided the
sample into 17~subgroups with 109~stars in each. The
approximated dependences $\sigma_{i}(t)$ and captions in the
plots in Fig.~5a show that the power-law indices for all
the semi-axes are appreciably different from those obtained
for all thin-disk stars in the solar neighborhood
(see below). In particular, the values of $\gamma$ for the 
major and middle semi-axes increased slightly, becoming
$0.26 \pm 0.04$ and $0.32 \pm 0.03$, respectively. At the same
time, the power-law index for the minor semi-axis
strongly decreased, becoming equal to zero within $3\sigma$
($\gamma_{3} = 0.07 \pm 0.03$); i.\,e., the velocity dispersion in the
direction perpendicular to the Galactic plane is essentially
independent of age. Recall that this index was
always appreciably larger, and not smaller, than the
other two semi-axes for the entire sample of nearby,
thin-disk stars (see, for example,~[1,~9,~13,~14]). The
power-law index for the age dependence of the total
residual velocity remained virtually the same: 
$\sigma_{V_{res}} =0.26\pm0.03$. (Recall that the power-law indices for the
age dependences of the semi-axes and total residual velocity
dispersion for all nearby, thin-disk stars were
$0.22\pm0.03$, $0.26\pm0.02$, $0.27\pm0.02$, and $0.24\pm0.02$,
respectively~[1]).

The ratio of the middle to the major semi-axes
in Fig~5b does not display appreciable variations
with age (the probability that the formal correlation
arose due to chance is $P \gg 5\,\%$), being, on average,
equal to $\langle \sigma_{2}/\sigma_{1}\rangle = 0.49 \pm 0.01$. 
This ratio is much smaller than in a stationary, rotating stellar
system, when $\sigma_{2}/\sigma_{2} = [-B/(A - B)]^{1/2}$; with 
$A = 13.7$ ~km\,s$^{-1}$~kpc and $B = -12.9$ ~km\,s$^{-1}$~kpc
the ''standard'' values of the Oort constants ($A = 15$
and $B = -10$ ~km\,s$^{-1}$~kpc), the ratio of the semi-axes,
0.58, remains larger than the value we have
obtained, beyond the errors. The ratios of the minor
to major semi-axes show a significant ($P < 1\,\%$) but
small decrease with age, since the dispersion of the $W$
velocity component depends only very weakly on age.
Moreover, this ratio became very small, 
$\langle\sigma_{3}/\sigma_{1}\rangle = 0.23 \pm 0.01$. 
Neither coordinate of the vertex deviation
displays appreciable correlations in Fig.~5c ($P > 5\,\%$); they 
both pass through zero somewhere in the
middle of the age range, and never take on values far
from zero, even for young ages. Their mean values are
$\langle L_{\odot} \rangle = 0.7^{\circ} \pm 0.5^{\circ}$ 
and $\langle B\rangle = 1.9^{\circ} \pm 1.1^{\circ}$.

The $U$ and $V$ components of the solar velocity
in Fig.~5d display appreciable correlations ($P \approx 1\,\%$).
The largest variations are observed for $U$; however, if
we exclude two extreme points with the largest ages
from the plot, the dependence essentially disappears,
suggesting that this dependence is most likely due
to selection effects associated with the wide range
in $R_{m}$. Neither of the other two components display
significant differences for their values at the opposite
ends of the age interval. As a result, the mean values
for $V$ for the stars in the subgroups of different ages
are roughly the same; i.\,e., their angular momentum
does not depend on age.
Due to the appreciable variations in $U$ and insignificant
variations in $W$, a significant slope for the
age dependences in Fig.~5e is observed only for the
longitude of the apex of the solar motion ($P \approx 1\,\%$),
but this is likely due to selection effects associated
with the comparatively wide range in $R_{m}$ we have
adopted. According to the data in Fig.~5e, the mean
coordinates of the apex of the solar motion are 
$\langle l_{\odot} \rangle = 70^{\circ} \pm 7^{\circ}$ 
and $\langle b_{\odot} \rangle = 41^{\circ} \pm 2^{\circ}$.

To test the validity of our conclusions about the
independence of metallicity of the parameters of the
velocity ellipsoids for stars born near the solar Galactocentric
distance, we divided sample described above
into 11~subgroups in $[Fe/H]$. As usual, we calculated
the velocity ellipsoids for each of these subgroups and
constructed the corresponding dependences. Indeed,
without exception, the velocity-ellipsoid parameters
displayed an absence ($P \geq 5\,\%$) of any dependence on
metallicity (the corresponding figure is not presented,
in view of the absence of positive information).

\section*{CONCLUSION}
Thus, excluding the influence of the radial migration
of stars (i.e., selection effects) led to appreciable
changes not only in the derived velocity-ellipsoid parameters
for thin-disk stars, but also their age dependences.
Our most significant result is the disappearance
of the difference in the angular momenta for stars
with different metallicities and ages. This means that,
within the thin disk, the angular momentum of stars
cannot be used as a statistical indicator of their ages.
For stars near the disk, this quantity is more likely
an indicator of the Galactocentric distance of their
birth places: the rotational velocity about the Galactic
center for disk stars near the solar Galactocentric distance
decreases with decreasing mean orbital radius.

The components of the solar velocity relative
to the LSR obtained via interpolation at the solar
Galactocentric distance, 
$(U_{\odot}, V_{\odot},W_{\odot})_{LSR} =(5.1\pm0.4, 7.9\pm0.5, 7.7\pm0.2)$
~km\,s$^{-1}$, differ slightly from
the usually used values. For example, the values
obtained in~[16] based on the proper motions of
14 000~nearby dwarfs from the Hipparcos catalog are
($10.0 \pm 0.4$, $5.3 \pm 0.6$, $7.2 \pm 0.4$)~km\,s$^{-1}$, 
while those obtained in~[17] using the data of the catalog~[13] 
by extrapolating the dispersions of the residual velocities
to zero are ($8.7\pm0.5$, $6.2\pm2.2$, $7.2\pm0.8$) ~km\,s$^{-1}$. 
If the initial assumption of the constancy of theme an orbital
radii of stars is valid, the method for determining the
motion of the Sun relative to the LSR used in the
current study should be more trustworthy and free of
systematic shifts, which are unavoidable in the case of
extrapolation or the use of samples containing stars of
diverse ages that were born at various distances from
the Galactic center.

The age dependences for the velocity ellipsoids
also changed appreciably. For example, the power-law
indices for the age dependences of the major,
middle, and minor semi-axes became equal to $0.26 \pm 0.04$, 
$0.32 \pm 0.03$, and $0.07 \pm 0.03$, respectively. Thus,
as before, the values of $\gamma_{1}$ and $\gamma_{2}$ can be explained
in a natural way by relaxation processes associated
with stochastic spiral density waves~[18]. The very
small value of the power-law index for the velocity
dispersion perpendicular to the Galactic plane suggests
that ''heating'' by spiral density waves alone
may be sufficient to explain this component as well.
Recall that, to explain the large value of this index
obtained earlier, we were led to invoke ''heating'' of
the stars by molecular clouds~[19], or even clusters of
dark matter from disrupted satellite galaxies under the
action of the tidal forces of our Galaxy~[20].
Thus, we have shown that, with increasing age,
the velocity ellipsoids for thin-disk stars born at the
solar Galactocentric distance increase only in the
plane of the disk, while they remain virtually constant
perpendicular to the disk. The shape of the ellipsoid
remains always far from equilibrium, and the direction
of its major semi-axis does not vary with age -- the
vertex deviation for the ellipsoid is constant and equal
to zero within the errors 
($\langle L \rangle = 0.7^{\circ} \pm 0.6^{\circ}$ 
and $\langle B \rangle = 1.9^{\circ} \pm 1.1^{\circ}$). 
Both coordinates of the apex of the solar
motion remain age independent within the errors,
having values $\langle l_{\odot} \rangle = 70^{\circ} \pm 7^{\circ}$ 
and $\langle b_{\odot} \rangle = 41^{\circ} \pm 2^{\circ}$.


\newpage

\begin{figure*}
\centering
\includegraphics[angle=0,width=0.80\textwidth,clip]{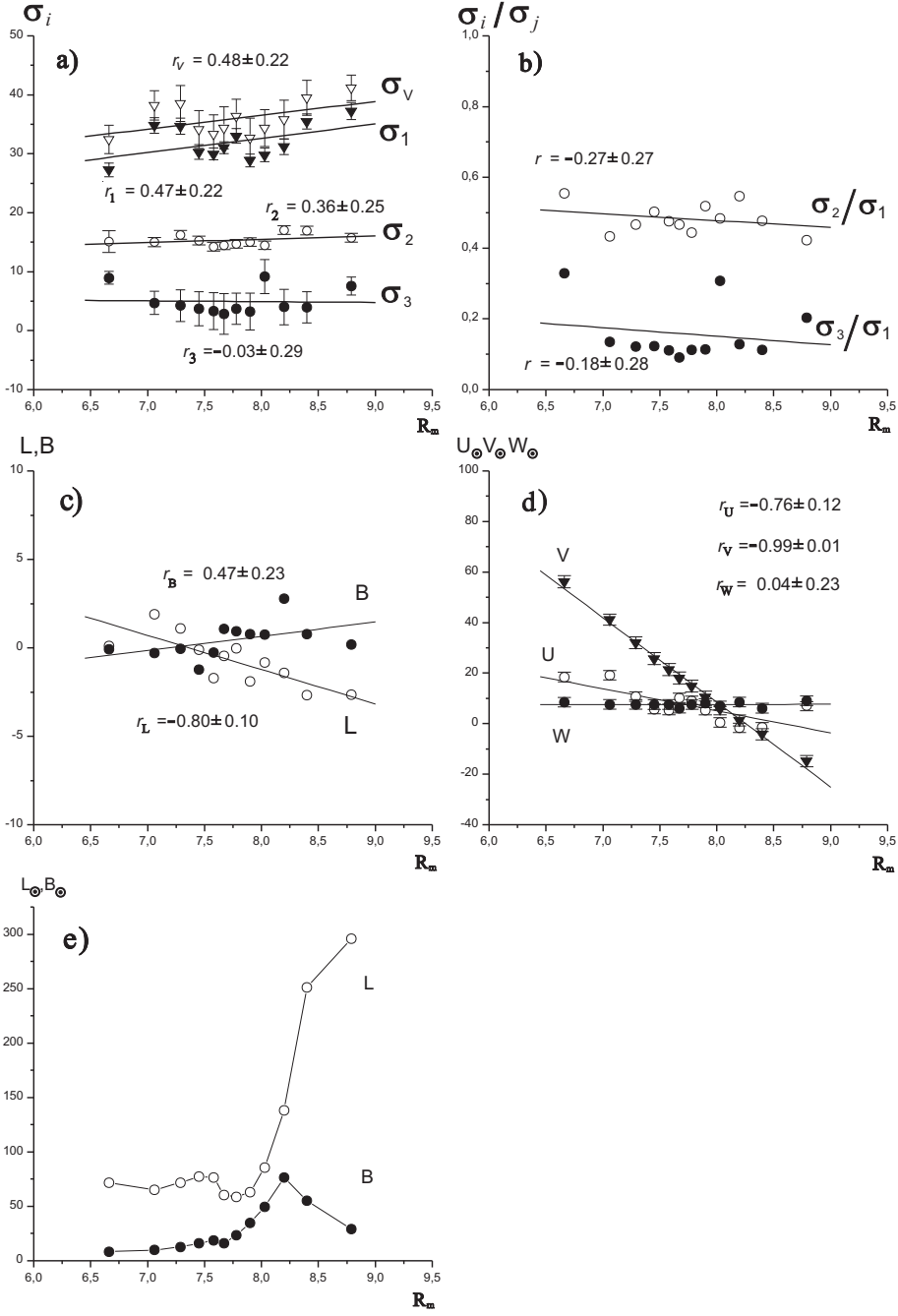}
\caption{Dependence on the mean orbital radius of (a) the semi-axes 
of the velocity ellipsoids and the total residual velocity,
(b) the ratios of the ellipsoid semi-axes, (c) the vertex coordinates, 
(d) the components of the solar velocity relative to the
corresponding centroids, and (e) the coordinates of the apex of the 
solar motion for nearby thin-disk stars. The solid lines are
regression fits, near which are indicated the correlation coefficients 
and their uncertainties. No regression fits are shown in (e)
due to the non-linear character of the dependences. Errors are 
presented everywhere, but in some cases, the formal parameter
errors are smaller than the symbols.}
\label{fig1}
\end{figure*}

\newpage

\begin{figure*}
\centering
\includegraphics[angle=0,width=0.90\textwidth,clip]{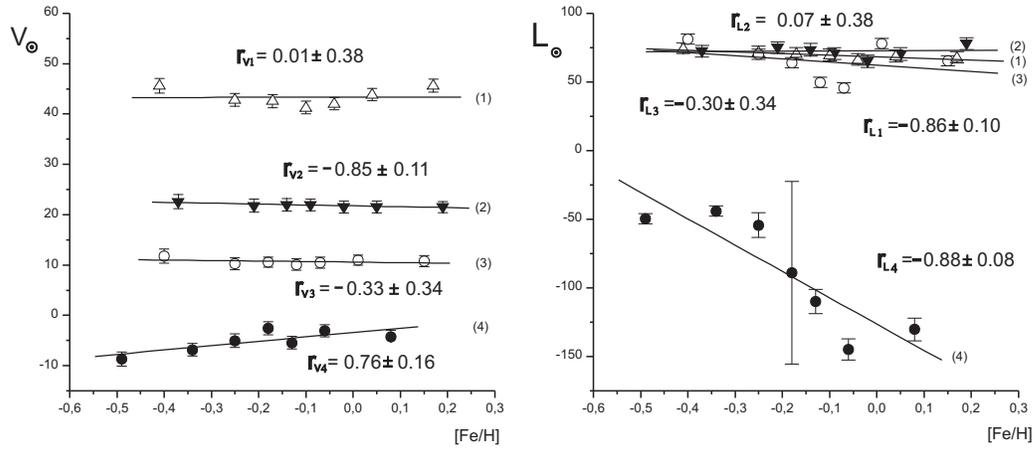}
\caption{Metallicity dependences for (a) the V component and (b) 
the longitude of the apex of the solar motion relative to
the corresponding centroids for four groups of thin-disk stars in 
narrow ranges of mean orbital radii: 1 $R_{m} < 7.37$kpc, 2
$7.37 < R_{m} < 7.72$kpc, 3 $7.72 < R_{m} < 8.10$kpc, 
4 $R_{m} > 8.10$kpc. Notation is the same as in Fig.~1}
\label{fig2}
\end{figure*}

\newpage

\begin{figure*}
\centering
\includegraphics[angle=90,width=0.96\textwidth,clip]{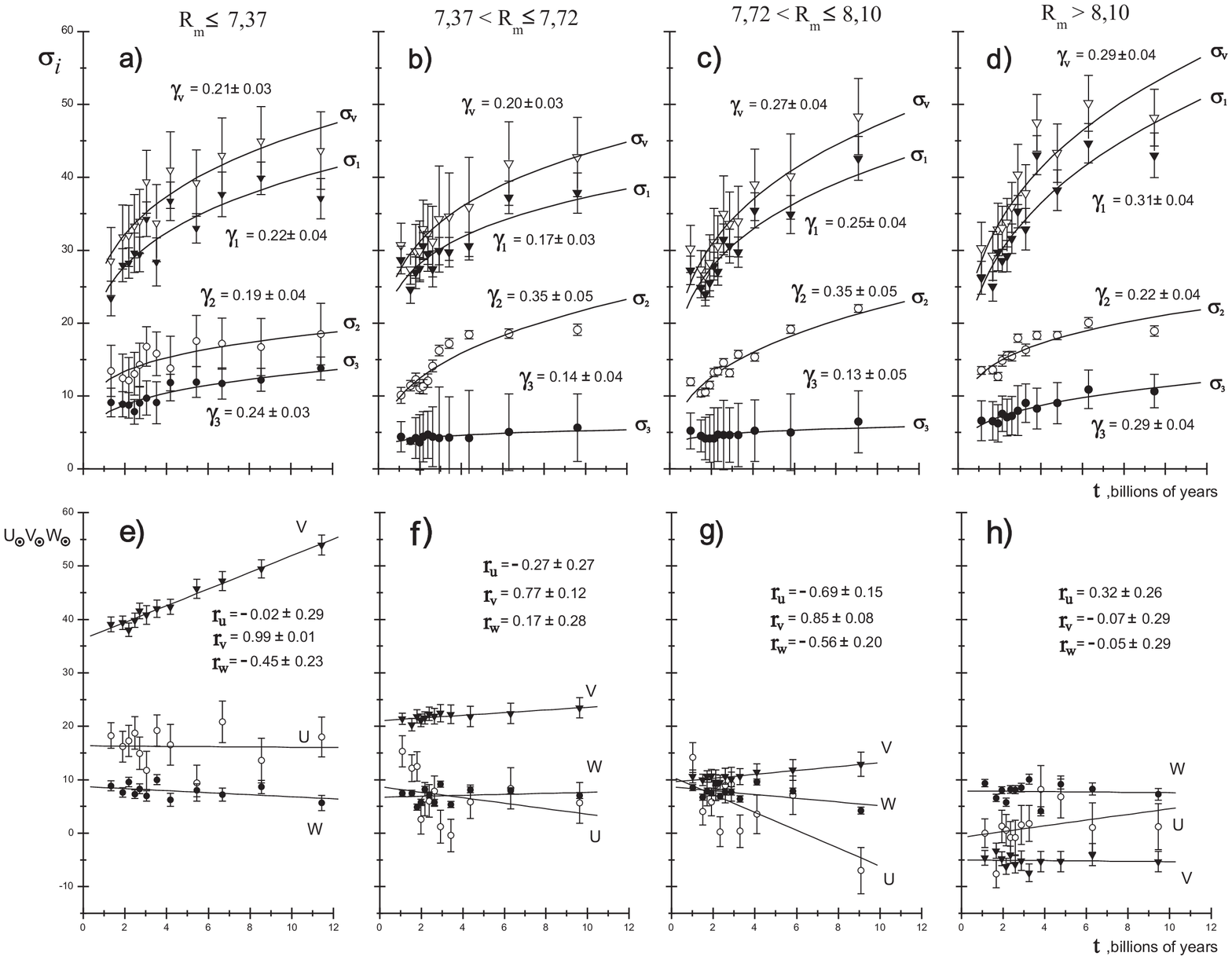}
\caption{Age dependences of the semi-axes of the velocity ellipsoids 
and the total residual velocity (upper row) and of the
components of the solar velocity relative to the corresponding 
centroids (lower row) for thin-disk stars within four intervals
of mean orbital radii (indicated above). The solid curves in the upper
 plots show power-law fits to the dependences; the power-law
indices and their errors are indicated. The solid curves in the lower 
plots show linear fits, and the correlation coefficients
and their errors are presented. The bars show the uncertainties in the 
corresponding quantities.}
\label{fig3}
\end{figure*}

\newpage

\begin{figure*}
\centering
\includegraphics[angle=90,width=0.96\textwidth,clip]{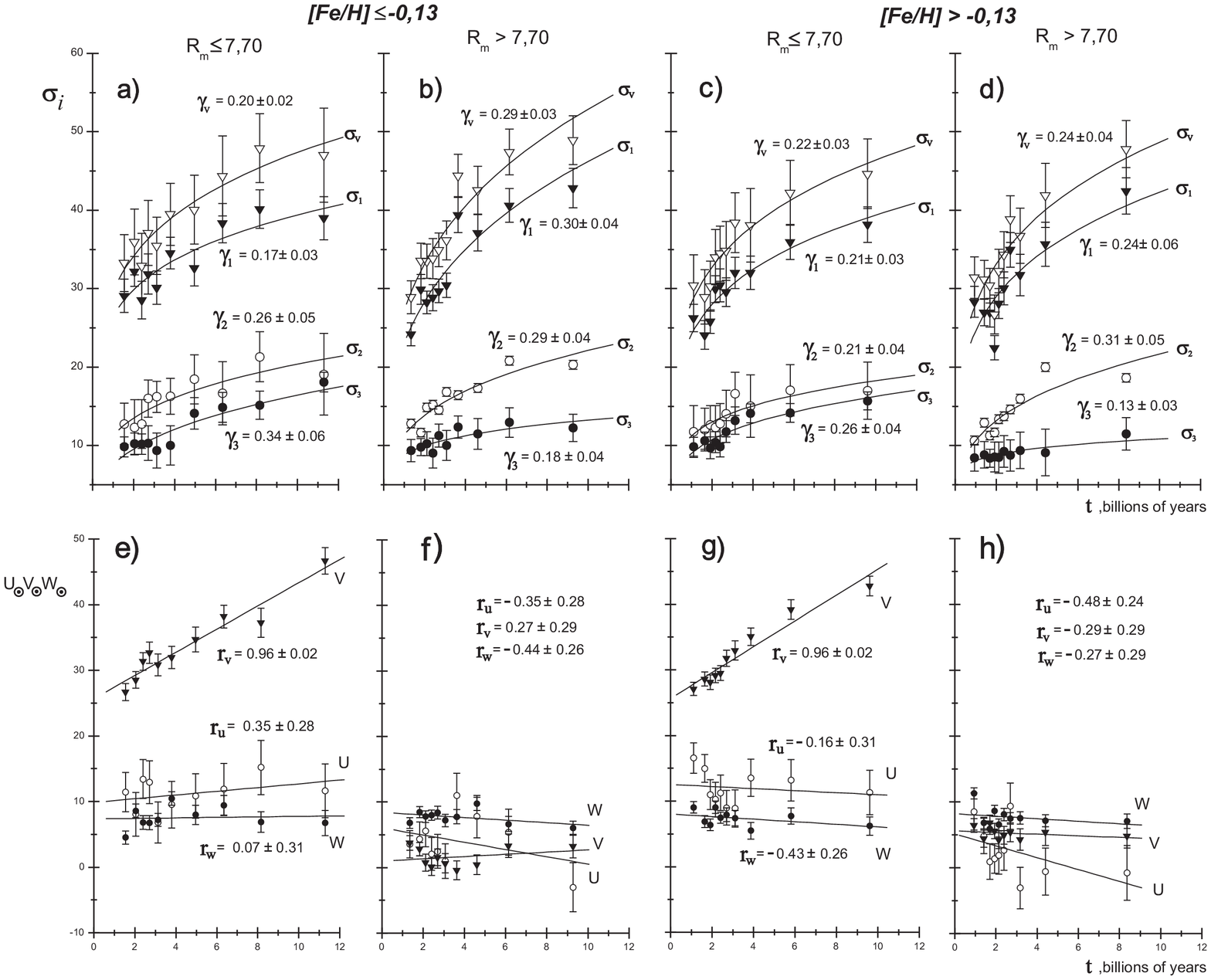}
\caption{Age dependences for the velocity ellipsoid semi-axes and the 
total residual velocity (upper row) and for the components
of the solar velocity relative to the corresponding centroids 
(second row) for thin-disk stars divided into two metallicity groups
(indicated above), and for each group divided into two subgroups in 
mean orbital radius (indicated above). The captions and
notation are as in Fig.~3.}
\label{fig4}
\end{figure*}

\newpage

\begin{figure*}
\centering
\includegraphics[angle=0,width=0.96\textwidth,clip]{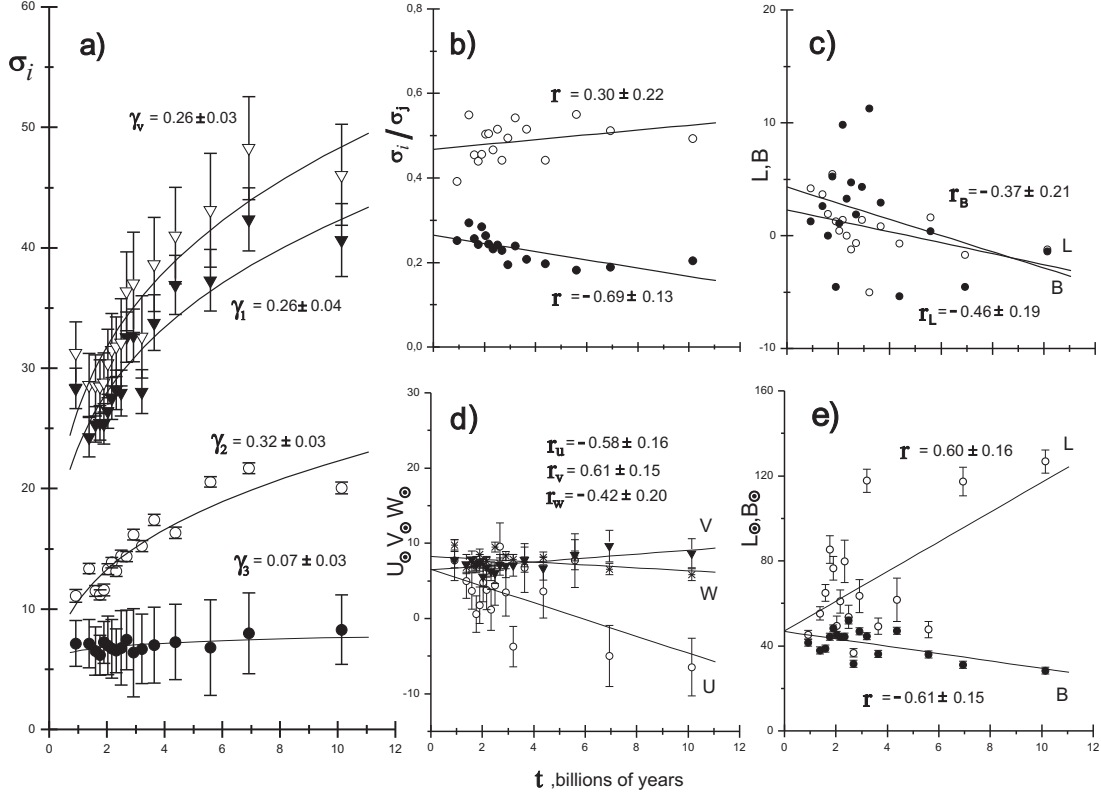}
\caption{Age dependences of the velocity ellipsoids for single F--G 
thin-disk stars with age uncertainties $\epsilon _{t} < \pm3$ billion
years, selected using our criterion and excluding stars in moving 
groups and in the narrow interval of mean orbital radii
$7.70 < R_{m} < 8.40$kpc. Presented are the age dependences for (a) the 
velocity-ellipsoid semi-axes and total residual
velocity, (b) the ratios of the ellipsoid semi-axes, (c) the vertex 
coordinates, (d) the solar velocity components relative to the
corresponding centroids, and (e) the coordinates of the apex of the 
solar motion. Notation and captions are as in Fig.~3.}
\label{fig5}
\end{figure*}

\end{document}